\documentclass[11pt]{amsart}
\usepackage[margin=1in]{geometry}		
\usepackage[parfill]{parskip}			
\usepackage{graphicx}
\usepackage{amssymb}
\usepackage{epstopdf}
\title{A Relativistic Symmetrical Interpretation of the Dirac Equation in (1+1) Dimensions}
\author{Michael B. Heaney\\3182 Stelling Drive\\Palo Alto, CA 94303\\mheaney@alum.mit.edu}
\date{25 October 2015}				
\begin{document}
\maketitle
\begin{abstract}
This paper presents a new Relativistic Symmetrical Interpretation (RSI) of the Dirac equation in (1+1)D which postulates: quantum mechanics is intrinsically time-symmetric, with no arrow of time; the fundamental objects of quantum mechanics are transitions; a transition is fully described by a complex transition amplitude density with specified initial and final boundary conditions; and transition amplitude densities never collapse. This RSI is compared to the Copenhagen Interpretation (CI) for the analysis of Einstein's bubble experiment with a spin-$\frac{1}{2}$ particle. This RSI can predict the future and retrodict the past, has no zitterbewegung, resolves some inconsistencies of the CI, and eliminates some of the conceptual problems of the CI.
\end{abstract}
\section{Introduction}
At the 1927 Solvay Congress, Einstein presented a thought-experiment to illustrate what he saw as a flaw in the Copenhagen Interpretation (CI) of quantum mechanics \cite{Valentini}. In his thought-experiment (later known as Einstein's bubble experiment) a single particle is released from a source, evolves freely for a time, then is captured by a detector. Let us assume the single particle is a massive spin-$\frac{1}{2}$ particle. The particle is localized at the source immediately before release, as shown in Figure 1(a). The CI says that, after release from the source, the particle's probability density evolves continuously and deterministically into a progressively delocalized distribution, up until immediately before capture at the detector, as shown in Figures 1(b) and 1(c). Immediately after capture, the particle is known to be localized at the detector (assumed here to be at the same location as the source), as shown in Figure 1(d). The CI says that, upon measurement at the detector, the probability density shown in Figure 1(c) undergoes an instantaneous, indeterministic, and irreversible collapse into the different probability density shown in Figure 1(d). Einstein's bubble pops. Einstein believed this collapse was unphysical, implying the CI was an incomplete theory. Einstein suggested that some additional mechanism, not included in the CI, was needed to make the wavefunction progressively relocalize to its final measured shape as it neared detection. An earlier paper \cite{HeaneyKGE} presented a new Relativistic Symmetrical Interpretation (RSI) of the Klein-Gordon equation that postulated no collapse and had a mechanism that made the wavefunction progressively relocalize to its final measured shape. Another earlier paper \cite{HeaneySE} extended the RSI to the Schr\"odinger equation. This paper will show how the RSI can be extended to the Dirac equation, and work out some of the implications.
 \begin{figure}[htbp]
\begin{center}
\includegraphics[width=6.5in]{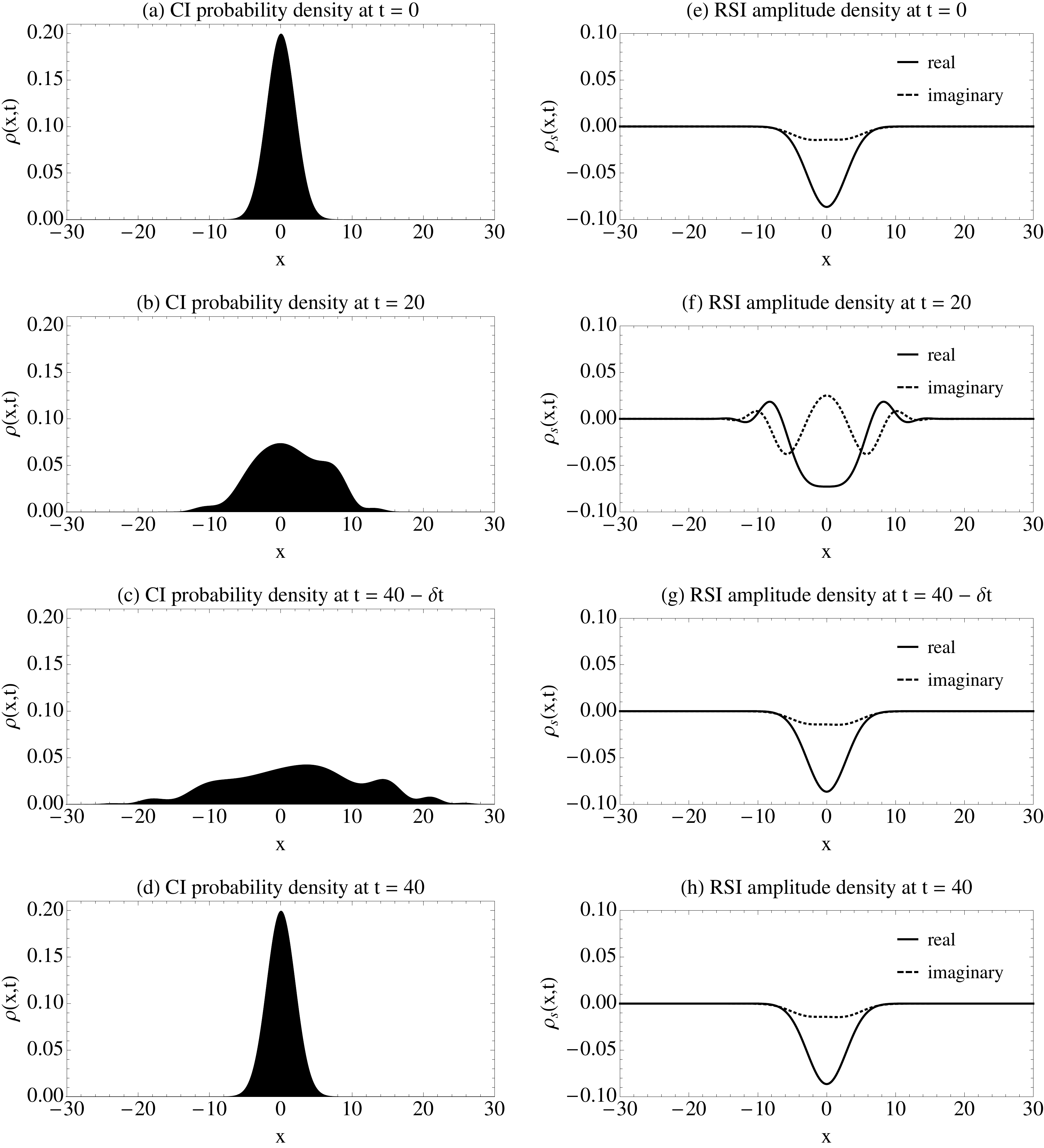}
\caption{\textbf{(a-d):} The Copenhagen Interpretation (CI) of the Dirac equation for a free particle, evolving from a gaussian probability density at $t=0$ to an identical probability density at $t=40$. The CI probability density $\psi^\dagger\psi$ shows strange distortions as it evolves. Upon measurement at $t=40$, the CI postulates that the wavefunction of (c) collapses instantaneously and irreversibly to the wavefunction of (d). \textbf{(e-h):} The Relativistic Symmetrical Interpretation (RSI) of the Dirac equation for a similar transition. The real and imaginary parts of the RSI transition amplitude density $\phi_+^\dagger\psi_+$ evolve smoothly and continuously at all times, with no strange distortions. The RSI has no wavefunction collapse. The RSI transition amplitude density $\psi_-^\dagger\phi_-$ gives the same results but with reversed phase, suggesting interpretation as the antiparticle.}
\label{default2}
\end{center}
\end{figure}
 \begin{figure}[htbp]
\begin{center}
\includegraphics[width=6.5in]{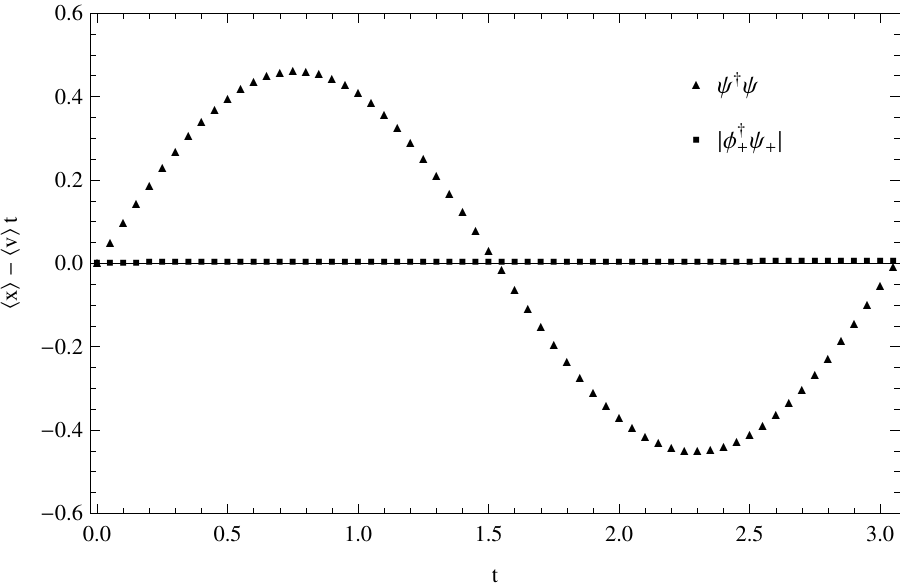}
\caption{The mean positions $\langle x \rangle$ of the CI probability density $\psi^\dagger\psi$ and of the absolute value of the RSI amplitude density $|\phi_+^\dagger\psi_+|$, with the drift motions $\langle v \rangle t$ subtracted out. The CI probability density shows the first cycle of zitterbewegung, while the RSI amplitude density shows no significant zitterbewegung. The small $(\pm 0.006)$ oscillations visible in the RSI amplitude density are present in the wavefunctions $\psi_+$ and $\phi_+^\dagger$, and decrease as the numerical accuracy of the calculations is increased, suggesting they are caused by numerical imprecision. The RSI transition amplitude density $\psi_-^\dagger\phi_-$ gives the same results.}
\label{default2}
\end{center}
\end{figure}
\section{The Copenhagen Interpretation of the Dirac Equation}
We will first describe how the CI of the Dirac equation (DE) explains Einstein's bubble experiment, using the CI postulates as presented in standard textbooks [3-10]. The CI implicitly assumes that quantum mechanics is a theory which describes an isolated, individual physical system, such as a massive spin-$\frac{1}{2}$ particle. The CI wavefunction postulate says this particle is completely described by a wavefunction $\psi(\vec{r},t)$, together with specified initial conditions. The CI implicitly assumes that we can arbitrarily specify the initial conditions of the particle's wavefunction. The CI evolution postulate says the wavefunction of a free, spin-$\frac{1}{2}$ particle of mass $m$ will evolve continuously and deterministically from these initial conditions according to the DE [3-10]:

\begin{equation}
\frac{1}{c}\sigma_0\frac{\partial\psi}{\partial t} + \sigma_x\frac{\partial\psi}{\partial x} + \frac{imc}{\hbar}\sigma_z\psi=0,
\label{ }
\end{equation}
where
\begin{equation}
\sigma_0 \equiv
\left(
\begin{array}{ccc}
1 & 0\\
0 & 1\\
\end{array}
\right),
\label{ }
\end{equation}
\begin{equation}
\sigma_x \equiv
\left(
\begin{array}{ccc}
0 & 1\\
1 & 0\\
\end{array}
\right),
\label{ }
\end{equation}
\begin{equation}
\sigma_z \equiv
\left(
\begin{array}{ccc}
1 & 0\\
0 & -1\\
\end{array}
\right),
\label{ }
\end{equation}
\begin{equation}
\psi \equiv
\left(
\begin{array}{c}
\psi_1 \\
\psi_2 \\
\end{array}
\right),
\label{ }
\end{equation}
$c$ is the speed of light, $m$ is the mass of the particle, and $2\pi\hbar$ is Planck's constant. We use the (1+1)D DE for computational tractability and ease of visualization. Appendix A shows how the (1+1)D DE can be obtained from the (3+1)D DE. The results of this paper should be extendable to the (3+1)D DE.

The DE has both negative frequency solutions ($\psi\sim e^{-i\omega t}$) and positive frequency solutions ($\psi\sim e^{+i\omega t}$). The CI assumes the standard energy operator $i\hbar\frac{\partial}{\partial t}$ to identify the negative frequency solutions as positive energy wavefunctions, and the positive frequency solutions as negative energy wavefunctions [3-10]. The CI implicitly assumes a complete set of solutions to the DE is required, which implies the general solution is a linear superposition of positive and negative energy wavefunctions, with specified initial conditions [3-10]. 

Angular momentum cannot be defined in one spatial dimension, so spin plays no role in the (1+1)D DE. The two-component spinor character of the Dirac wavefunction in (1+1)D is related only to the fact that the energy can be either positive or negative. 

The Hermitian conjugate of the DE is also a valid wave equation:

\begin{equation}
\frac{1}{c}\frac{\partial \psi^\dagger}{\partial t}\sigma_0 + \frac{\partial \psi^\dagger}{\partial x}\sigma_x - \frac{imc}{\hbar}\psi^\dagger\sigma_z=0.
\label{ }
\end{equation}

If we multiply the DE on the left by $\psi^\dagger$, multiply the Hermitian conjugate of the DE on the right by $\psi$, and add the two resulting equations, we get a CI local conservation law:

\begin{equation}
\frac{\partial\rho}{\partial t}+\nabla\cdot j=0,
\label{ }
\end{equation}
where
\begin{equation}
\rho(x,t)\equiv\psi^\dagger\psi,
\label{ }
\end{equation}
and 
\begin{equation}
j(x,t)\equiv c\psi^\dagger\sigma_x\psi.
\label{ }
\end{equation}

The CI interprets $\rho(x,t)$ as the probability density for finding the particle at position $x$ at time $t$, and $j(x,t)$ as the probability density current. The fact that $\rho(x,t)$ is always real and positive is taken as a confirmation of this interpretation. 

If $\psi(x,t)$ is normalized and goes to zero at $x=\pm\infty$, we can integrate the CI local conservation law over all space to get a CI global conservation law:

\begin{equation}
\int_{-\infty}^{+\infty}\rho(x,t)dx = 1,
\label{ }
\end{equation}

which means the probability of finding the particle somewhere in space is conserved and equal to one. 

Figure 1(a) shows the CI probability density at the initial time $t_i\equiv 0$ for the normalized gaussian wavefunction given by:

\begin{equation}
\psi(x,0) \equiv
\left(
\frac{1}{32\pi}
\right)^{1/4}e^{-\frac{x^2}{16}}
\left(
\begin{array}{c}
1 \\
1 \\
\end{array}
\right),
\label{ }
\end{equation}

where we use natural units, set $m=1$, and assume a standard deviation of 2 \cite{Thaller2}. 

The CI implicitly assumes that we can arbitrarily specify the initial wavefunction, while the RSI explicitly assumes we cannot. This can cause qualitatively different behavior (zitterbewegung) between the CI and the RSI, which helps illuminate one of the ways the two theories differ. For this reason, we have chosen a CI initial wavefunction that shows this difference. This CI initial wavefunction is a linear superposition of positive and negative energy wavefunctions. Figures 1(a-c) show the evolution of the probability density according to the CI of the DE. In 1930 Schr\"odinger discovered that the CI of the DE predicts a rapid oscillating motion of the mean position of a free spin-$\frac{1}{2}$ particle in vacuum, naming it zitterbewegung \cite{Schroedinger}. He showed that zitterbewegung is caused by interference between the positive and negative energy components of the wavefunction. Figure 2 shows the first cycle of this zitterbewegung. The velocity of these oscillations varies between $\pm c$. 

Let us assume, for simplicity, that a measurement of the location of the particle at the final time $t_f\equiv 40$ gives the same wavefunction and probability density as at $t_i\equiv 0$, as shown in Figure 1(d). The CI measurement postulate assumes the transition amplitude $A$ for a particle having the wavefunction $\psi(x,t_f-\delta t)$ to be found having the final wavefunction $\phi(x,t_f)$ is:

\begin{equation}
A\equiv\int_{-\infty}^{+\infty}\phi^\dagger(x,t_f) \psi(x,t_f-\delta t)dx.
\label{ }
\end{equation}

Note that the integral is evaluated at the time $t_f$. For the transition from Figure 1(c) to Figure 1(d), numerical calculations give $A = -0.584 - 0.010 i$. The CI measurement postulate also assumes the transition probability $P$ for such a transition is:

\begin{equation}
P\equiv A^\ast A.
\label{ }
\end{equation}

For the transition from Figure 1(c) to Figure 1(d), numerical calculations give $P=0.341$.

The CI collapse postulate assumes that, upon measurement at $t_f$, the wavefunction $\psi(x,t_f-\delta t)$, and therefore the CI probability density of Figure 1(c), undergoes an instantaneous, indeterministic, and time-asymmetric collapse into the CI probability density of Figure 1(d). The CI postulates that the collapsed wavefunction $\phi(x,t_f)$ of Figure 1(d) will then evolve continuously and deterministically according to the DE, until the next measurement. 
\section{The Relativistic Symmetrical Interpretation of the Dirac Equation}
Let us review the time-asymmetric postulates of the CI, and attempt to construct a relativistic symmetrical interpretation (RSI). The CI explicitly or implicitly postulates: quantum mechanics has an intrinsic arrow of time, is asymmetrical in time, and can predict the future but not retrodict the past; quantum mechanics is a theory which describes an isolated, individual physical system such as a particle; this physical system is completely described by a wavefunction $\psi(\vec{r},t)$ or a vector in Hilbert space, together with specified initial conditions; there are many different types of experimental observables; we can arbitrarily specify the initial conditions; the wavefunction $\psi(\vec{r},t)$ evolves continuously and deterministically from the initial conditions towards the final conditions according to the appropriate wave equation until it is measured; upon measurement the evolved wavefunction $\psi(\vec{r},t_f)$ collapses discontinuously, indeterministically, and irreversibly into a different wavefunction $\phi(\vec{r},t_i)$; and after measurement this different wavefunction evolves continuously and deterministically from the initial conditions $\phi(\vec{r},t_i)$ according to the appropriate wave equation, until the next measurement.

Now let us try to replace each of these CI postulates with tentative RSI postulates: quantum mechanics has no intrinsic arrow of time, is symmetrical in time, and can predict the future and retrodict the past; the fundamental objects of quantum mechanics are transitions; a transition is completely described by a complex transition amplitude density together with specified initial and final conditions; transition probabilities are the only experimental observables; transition amplitude densities and wavefunctions never collapse; and after measurement (assumed to be minimally disturbing) transition amplitude densities and wavefunctions continue to evolve continuously and deterministically according to the appropriate wave equations. 

In the 1940's St\"uckelberg \cite{Stuckelberg} and Feynman \cite{Feynman} reconsidered quantum mechanics from a space-time viewpoint. From this viewpoint, they interpreted the positive energy solutions of the DE as positive energy electrons moving forwards in time, and the negative energy solutions as negative energy electrons moving backwards in time. They also showed that negative energy electrons moving backwards in time could be interpreted as positive energy positrons moving forwards in time. This is now known as the St\"uckelberg-Feynman interpretation, and their space-time viewpoint is now known as the block-universe viewpoint. The St\"uckelberg-Feynman interpretation was later shown to be a direct consequence of the CPT theorem, and is described in most standard textbooks [3-10]. We will adopt and extend the St\"uckelberg-Feynman interpretation to the RSI. Extending the St\"uckelberg-Feynman interpretation to the RSI suggests interpreting positive energy wavefunctions $\psi_+(\vec{r},t)$ as waves that satisfy only the initial conditions and move forwards in time, and negative energy wavefunctions $\psi_-(\vec{r},t)$ as waves that satisfy only the final conditions and move backwards in time. We will further postulate that $\phi^\dagger_+(\vec{r},t)\psi_+(\vec{r},t)$ is the RSI transition amplitude density for a positive energy particle to go forwards in time from the specified initial condition $\psi_+(\vec{r},t_i)$ to the specified final condition $\phi_+(\vec{r},t_f)$. Then $\psi^\dagger_-(\vec{r},t)\phi_-(\vec{r},t)$ is the RSI transition amplitude density for a negative energy particle to go backwards in time from the specified final condition $\phi_-(\vec{r},t_f)$ to the specified initial condition $\psi_-(\vec{r},t_i)$, which will look like a positive energy antiparticle going forwards in time from the initial condition to the final condition. These postulates also imply that neither the initial nor final conditions can be arbitrarily specified: they must be either both positive energy or both negative energy. This is qualitatively different than the CI, which postulates that the initial conditions can be arbitrarily specified, which generally requires an initial state which is a superposition of a positive energy wavefunction and a negative energy wavefunction. In the CI, this superposition satisfies only the initial conditions, and only goes forwards in time. Even if the CI were to restrict the initial wavefunction to be only a positive energy wavefunction, the resulting behavior would still be significantly different than the RSI. This was described in detail in an earlier paper \cite{HeaneyKGE}.

Let us work out the RSI positive energy case. Assume a positive energy wavefunction $\psi_+(x,t)$ which satisfies only the initial conditions $\psi_+(x,t_i)$ and obeys the DE:

\begin{equation}
\frac{1}{c}\sigma_0\frac{\partial\psi_+}{\partial t} + \sigma_x\frac{\partial\psi_+}{\partial x} + \frac{imc}{\hbar}\sigma_z\psi_+=0.
\label{ }
\end{equation}

Now assume a second, different positive energy wavefunction $\phi_+(x,t)$ which satisfies only the final conditions $\phi_+(x,t_f)$ and obeys the DE:

\begin{equation}
\frac{1}{c}\sigma_0\frac{\partial\phi_+}{\partial t} + \sigma_x\frac{\partial\phi_+}{\partial x} + \frac{imc}{\hbar}\sigma_z\phi_+=0.
\label{ }
\end{equation}

Let us take the Hermitian conjugate of this equation:

\begin{equation}
\frac{1}{c}\frac{\partial \phi_+^\dagger}{\partial t}\sigma_0 + \frac{\partial \phi_+^\dagger}{\partial x}\sigma_x - \frac{imc}{\hbar}\phi_+^\dagger\sigma_z=0.
\label{ }
\end{equation}

If we multiply Equation 14 on the left by $\phi_+^\dagger$, and multiply Equation 16 on the right by $\psi_+$, and then add the two resulting equations, we get a local conservation law: 

\begin{equation}
\frac{\partial\rho_s}{\partial t}+\nabla\cdot j_s=0,
\label{ }
\end{equation}
where
\begin{equation}
\rho_s(x,t)\equiv\phi_+^\dagger\psi_+,
\label{ }
\end{equation}
and 
\begin{equation}
j_s(x,t)\equiv c\phi_+^\dagger\sigma_x\psi_+.
\label{ }
\end{equation}

The RSI interprets $\rho_s(x,t)$ as the symmetrical transition amplitude density, and $j_s(x,t)$ as the symmetrical transition amplitude density current. These are generally complex functions, and are not interpreted as probability densities. The RSI symmetrical transition amplitude $A_s$ is defined as the integral over all space of the RSI transition amplitude density:

\begin{equation}
A_s\equiv \int_{-\infty}^{+\infty}\phi_+^\dagger(x,t)\psi_+(x,t)dx.
\label{ }
\end{equation}

Note that the integral can be evaluated at any time $t$. $A_s$ is a complex constant, independent of the time $t$ and the position $x$. $A_s$ is the amplitude that a particle found in the initial state $\psi_+(x,t_i)$ will later be found in the final state $\phi_+(x,t_f)$. The RSI symmetrical transition probability $P_s$ is defined as $P_s\equiv A_s^\ast A_s$. $P_s$ is the probability that a particle found in the initial state $\psi_+(x,t_i)$ will later be found in the final state $\phi_+(x,t_f)$. Note that $P_s$ is always a positive real constant.

Given the same initial and final wavefunctions, the RSI gives the same results for the transition amplitude as the CI. This is a direct consequence of the unitarity of quantum mechanics: the evolution of any initial wavefunction is given by $\vert\Psi(t)\rangle=U(t-t_i)\vert\Psi(t_i)\rangle$, where $U(t-t_i)$ is a unitary time evolution operator. Similarly, the evolution of any final wavefunction is given by $\vert\Phi(t)\rangle=U(t-t_f)\vert\Phi(t_f)\rangle$. The RSI transition amplitude between these two states at any time $t$ is defined to be $\langle\Phi(t)\vert\Psi(t)\rangle=\langle\Phi(t_f)\vert U^\dag(t-t_f)U(t-t_i)\vert\Psi(t_i)\rangle=\langle\Phi(t_f)\vert U(t_f-t)U(t-t_i)\vert\Psi(t_i)\rangle=\langle\Phi(t_f)\vert U(t_f-t_i)\vert\Psi(t_i)\rangle=\langle\Phi(t_f)\vert\Psi(t_f)\rangle$, which is equal to the CI transition amplitude between the same two states.

However, the RSI gives a qualitatively different interpretation for what is happening: the CI postulates that the wavefunction $\psi_+(x,t)$ evolves continuously and smoothly from time $t_i$ to time $t_f$, then at time $t_f$ collapses discontinuously and irreversibly into the wavefunction $\phi_+(x,t_f)$, similar to Figures 1(a-d). This is why the CI specifies that the transition amplitude of Equation 12 is to be evaluated only at the time of collapse $t_f$. The RSI postulates that $\rho_s(x,t)$ evolves continuously and smoothly from time $t_i$ to time $t_f$ and beyond, with no collapse, as shown in Figures 1(e-h). This is why the RSI allows the symmetrical transition amplitude of Equation 20 to be evaluated at any time.

Now consider the specific case where $\psi_+(x,t_i)$ and $\phi_+(x,t_f)$ are both equal to the positive energy part of the gaussian wavefunction given by Equation 11. Figures 1(e-h) show the evolution of the real and imaginary parts of the resulting RSI transition amplitude density $\rho_s(x,t)$. The RSI transition amplitude density delocalizes between $t_i=0$ and $t=20$, then relocalizes between $t=20$ and $t_f=40$. The transition is smooth and continuous, with neither strange distortions nor abrupt collapses. Also, the mean position of the absolute value of $\rho_s(x,t)$ does not exhibit zitterbewegung, as shown in Figure 2. The small $(\pm 0.006)$ oscillations visible in $\rho_s(x,t)$ are present in both $\psi_+(x,t)$ and $\phi_+^\dagger(x,t)$, and decrease as the numerical accuracy is increased, suggesting they are due to numerical imprecision of the calculations. For the transition from Figure 1(e) to Figure 1(h), numerical calculations give $A_s = -0.607 - 0.161 i$ and $P_s=0.394$. Note that these differ from the CI values calculated earlier because the RSI wavefunctions are only the positive energy parts of the CI wavefunctions. The negative energy parts of the same CI wavefunctions give the same results for the transition probability, but with the transition amplitude phase reversed, suggesting interpretation as the antiparticle. 
\section{Discussion} 
The RSI is a new type of Symmetrical Interpretation (SI) of quantum mechanics. SI's of quantum mechanics have a long history, dating back at least to a 1921 paper by Schottky \cite{Schottky}. Many different types of SI's have been developed over the past century [15-43]. The similarities and differences between the RSI and other types of SI have been described elsewhere \cite{HeaneyKGE,HeaneyDC,HeaneySE}.

In 1932 Dirac showed that all the experimental predictions of the CI of quantum mechanics can be formulated in terms of transition probabilities \cite{Dirac2}. The RSI inverts this fact by postulating that quantum mechanics is a theory which experimentally predicts \textit{only} transition probabilities. This implies the RSI has the same predictive power as the CI. But this does not imply the RSI makes all of the same predictions as the CI. 

For example, the CI of the DE predicts zitterbewegung, as shown above. Zitterbewegung of a free electron has a frequency of order $10^{21}$ Hz and an amplitude of order $10^{-13}$ m, which is beyond measurement with current technology. It is natural that wavefunctions oscillate, but zitterbewegung is a qualitatively different kind of oscillation, where the center of mass of a free wavefunction oscillates back and forth in space. This violates the conservation of energy-momentum, which seems unphysical. Why would a single electron in free space, with no forces acting on it, spontaneously move back and forth at the speed of light? Some physicists believe zitterbewegung is not a real physical phenomenon, but just an artifact of an incorrect interpretation of the DE \cite{Itzykson}. Other physicists believe zitterbewegung is a real physical phenomenon, with experimentally observable effects \cite{Sakurai}. Zitterbewegung of a free particle is an unresolved puzzle in the foundations of the CI of quantum mechanics. The RSI of the DE presented here has no zitterbewegung, resolving this puzzle. However, if the positive and negative energy wavefunction components of the RSI transition amplitudes given above are Lorentz-transformed into a different inertial frame, then the Lorentz transformations will change each wavefunction into a superposition of positive and negative energy wavefunctions, producing zitterbewegung. This may indicate a serious flaw of the RSI. Alternatively, note that the CI of quantum mechanics assumes an arrow of time and assumes particles are the fundamental building blocks of nature. The conventional interpretation of the special theory of relativity makes these same assumptions, and also assumes that energy is always positive. But the RSI of quantum mechanics makes the opposite assumptions. This suggests the special theory of relativity may need reinterpretation or extension to be fully compatible with the RSI.

How can the RSI prediction of no zitterbewegung be reconciled with the experimental simulation of zitterbewegung for a particle \cite{Gerritsma2}? This simulation was done with a calcium ion in a Paul trap, and was actually a simulation of the CI of the DE, which is known to predict zitterbewegung. An experimental simulation of the RSI of the DE would not show zitterbewegung. How can the RSI be reconciled with the direct observation of apparent zitterbewegung in solid state systems, such as electron channeling in a crystal \cite{Catillon, Hestenes}? For an electron in a crystal, the apparent zitterbewegung is actually the oscillatory motion of the electron traveling through a periodic lattice potential, and is a completely different phenomenon than the zitterbewegung of a free spin-$\frac{1}{2}$ particle in vacuum \cite{Zawadzki2}. 

The explicit and implicit postulates of the CI have several asymmetries in time: only the initial conditions of the wavefunction are specified, the wavefunction is evolved only forward in time, the transition amplitude is calculated only at the time of measurement, wavefunction collapse happens only at the time of measurement, and wavefunction collapse happens only forwards in time. This seems unphysical: shouldn't the fundamental laws of nature be time-symmetric? Consider the details of a specific example: according to the CI, the CI transition amplitude $A$ of Equation 12 must be evaluated only at the time of collapse. In contrast, according to the RSI, the RSI transition amplitude $A_s$ of Equation 20 can be evaluated at any time. But the RSI transition amplitude still gives exactly the same results as the CI transition amplitude! The fact that the transition amplitude need not be evaluated at a special time shows that quantum mechanics has more intrinsic symmetry than allowed by the CI postulates. Heisenberg \cite{Heisenberg} said \textquotedblleft Since the symmetry properties always constitute the most essential features of a theory, it is difficult to see what would be gained by omitting them in the corresponding language." The intrinsic time symmetry of a quantum transition amplitude is represented in the RSI postulates, but not in the CI postulates.

More generally, the CI implicitly assumes that quantum mechanics is only a predictive theory: first specify the initial wavefunction, then use the appropriate wave equation to evolve the wavefunction forwards in time, then make a measurement which irreversibly collapses the evolved wavefunction into a different wavefunction. As Dyson pointed out \cite{Dyson}, \textquotedblleft ...statements about the past cannot in general be made in quantum-mechanical language. For example, we can describe a uranium nucleus by a wave function including an outgoing alpha particle wave which determines the probability that the nucleus will decay tomorrow. But we cannot describe by means of a wave function the statement, \textquoteleft This nucleus decayed yesterday at 9 a.m. Greenwich time.'" Feynman also believed that quantum mechanics could not account for history \cite{Bernstein}.

When the CI is used retrodictively, attempting to determine what happened in the past given the present wavefunction, it usually does not work. Penrose \cite{Penrose} used an interferometer thought-experiment to show that using the CI retrodictively gives us \textquotedblleft ...completely the wrong answer!" Peres \cite{Peres} used a thought-experiment with an entangled pair of spin-$\frac{1}{2}$ particles to show that using the CI retrodictively results in paradoxes. Hartle \cite{Hartle} proved that in the CI \textquotedblleft ...correct probabilities for the past cannot generally be constructed simply by running the Schr\"odinger equation backwards in time from the present state."

The inability of the CI of quantum mechanics to describe or retrodict the past seems like a serious shortcoming for a theory which claims to be our best description of nature! Since the RSI is time-symmetric, it describes the future and past equally well, and makes correct predictions and retrodictions. For example, consider the experiments shown in Figure 1. Given the wavefunction of Figure 1(a), the CI can correctly predict the later wavefunctions of Figure 1(b) and 1(c). But given the wavefunction of Figure 1(d), the CI cannot retrodict the wavefunctions of Figures 1(a,b,c). In contrast, given any one of the transition amplitude densities of Figures 1(e,f,g,h), the RSI can predict and retrodict all of the other transition amplitude densities. 

The RSI explicitly postulates that the initial conditions must be either purely positive or purely negative energy. This severely limits the possible initial conditions. In contrast, the CI implicitly postulates that the initial conditions can be arbitrarily specified. This is very reasonable for the CI of nonrelativistic quantum mechanics, and agrees with all the nonrelativistic experimental evidence. All of the solutions to the Schr\"odinger equation have positive energy, and any arbitrarily specified initial wavefunction will have positive energy. It was reasonable to assume that this implicit postulate could be extended to relativistic quantum mechanics. However, every relativistic wave equation has both positive and negative energy solutions. This means that an arbitrarily specified initial condition will generally be a wavefunction which is a superposition of a positive energy component and a negative energy component. The CI later postulates that a positive energy wavefunction represents a particle, while a negative energy wavefunction represents an antiparticle. In the low energy limit, where particle number is constant, this implies the existence of a single particle which is a superposition of an electron and a positron. Such a particle has never been observed, casting doubt on the correctness of the CI postulate that the initial conditions can be arbitrarily specified. The RSI postulates that the initial wavefunction $\psi$ must be either positive energy or negative energy, and the sign of energy of the final wavefunction $\phi$ must match the sign of energy of $\psi$. This implies that a single particle which is a superposition of an electron and a positron will never occur, in agreement with all experimental results. This is analogous to the CI symmetrization postulate, which says the initial and final conditions of a group of identical particles cannot be arbitrarily specified.

The CI postulates two contradictory types of physical process: continuous, deterministic wavefunction evolution determined by a wave equation; and discontinuous, indeterministic wavefunction collapse upon measurement, determined by no known equation. Also, wavefunction evolution is Lorentz-invariant, while wavefunction collapse is not. We would expect the fundamental laws of nature to be Lorentz-invariant. This is part of the \textquotedblleft measurement problem" of the CI \cite{Wheeler}, and gives rise to many of the unresolved conceptual questions of the CI. The measurement problem has plagued the CI of quantum mechanics for close to a century now, and is considered a major unsolved problem of modern physics \cite{Smolin}. Since the RSI postulates only continuous, deterministic wavefunction and transition amplitude density evolution, it does not have at least this part of the measurement problem.

The CI of the DE predicts both positive and negative energy states, which makes all spin-$\frac{1}{2}$ particles unstable to decay and implies spin-$\frac{1}{2}$ particles can have negative energy, both contrary to experimental evidence. The CI claims to solve these problems by postulating infinite Dirac seas filled with negative energy particles, and reinterpreting the negative energy states as holes in the Dirac sea which behave as positive energy antiparticles. This explanation seems dubious: it might work for fermions, but it does not work for bosons. In contrast, the RSI does not identify a particle with a state, but instead identifies the transition of a particle with an amplitude density. This eliminates the instability and negative energy problems, and requires no infinite Dirac seas. The RSI predicts both particles and antiparticles as a direct consequence of the two possible types of transition amplitudes. The RSI explanation also works equally well for fermions and bosons. 

The Klein-Gordon equation (KGE) for spin-$0$ particles was discovered before the DE. The CI of the KGE predicts both positive and negative probability densities, which seems unphysical [2-10]. For this reason, the KGE was believed to be an invalid wave equation. The DE, which predicts only positive probability densities, was accepted as a valid wave equation. Then in 1934, Pauli and Weisskopf proposed a reinterpretation of the KGE: the probability densities could be multiplied by the particle's charge and reinterpreted as charge densities \cite{Pauli}. The existence of spin-$0$ particles with both positive and negative charges was taken as a confirmation of this interpretation. But this makes the CI interpretations of the DE and the KGE inconsistent: why does an electron have a probability density, but a Higgs boson does not? In contrast, the RSI has a consistent interpretation of both the DE and the KGE: they both have symmetrical transition amplitude densities, which are generally complex. But the predicted transition probabilities for both equations are always positive real constants. In the RSI, neither the DE transition amplitude density nor the KGE transition amplitude density needs to be multiplied by charge to give correct predictions for transition probabilities. 

Finally, the longstanding conceptual problems in the foundations of the CI of quantum mechanics suggest that something in the CI is fundamentally wrong. This paper suggests what is wrong: the CI implicitly assumes a quantum arrow of time, and implicitly assumes that the fundamental objects are wavefunctions (or vectors in Hilbert space) representing particles. It is ingrained in human experience and intuition that nature has an intrinsic arrow of time and is composed of particles, which leads us to implicitly extrapolate these concepts to the quantum level. These incorrect extrapolations are the cause of many conceptual problems in the CI of quantum mechanics. The RSI does not make these assumptions, and does not have the associated conceptual problems.
\section{Acknowledgements} 
The author thanks Eugene D. Commins for many useful conversations.
\appendix
\section{reduction of the (3+1) DE to the (1+1) DE}
The (3+1)D DE is [7-10]:

\begin{equation}
\frac{1}{c}\frac{\partial\psi}{\partial t}+\vec{\alpha}\cdot\nabla\psi+\frac{imc}{\hbar}\beta\psi=0,
\label{ }
\end{equation}

where\\
\[
\alpha_i\equiv
\left(
\begin{array}{ccc}
0 & \sigma_i\\
\sigma_i & 0\\
\end{array}
\right),
\]

\[
\sigma_x \equiv
\left(
\begin{array}{ccc}
0 & 1\\
1 & 0\\
\end{array}
\right),
\]

\[
\sigma_y \equiv
\left(
\begin{array}{ccc}
0 & -i\\
i & 0\\
\end{array}
\right),
\]

\[
\sigma_z \equiv
\left(
\begin{array}{ccc}
1 & 0\\
0 & -1\\
\end{array}
\right),
\]

\[
\beta\equiv
\left(
\begin{array}{ccc}
I & 0\\
0 & -I\\
\end{array}
\right),
\]

\[
I\equiv
\left(
\begin{array}{ccc}
1& 0\\
0 & 1\\
\end{array}
\right).
\]

If we assume $\psi$ is a function of $x$ and $t$ only, then the (3+1)D DE reduces to:\\

\[\frac{1}{c}\frac{\partial\psi_1}{\partial t} + \frac{\partial\psi_4}{\partial x} + \frac{imc}{\hbar}\psi_1 = 0,\]

\[\frac{1}{c}\frac{\partial\psi_2}{\partial t} + \frac{\partial\psi_3}{\partial x} + \frac{imc}{\hbar}\psi_2 = 0,\]

\[\frac{1}{c}\frac{\partial\psi_3}{\partial t} + \frac{\partial\psi_2}{\partial x} - \frac{imc}{\hbar}\psi_3 = 0,\]

\[\frac{1}{c}\frac{\partial\psi_4}{\partial t} + \frac{\partial\psi_1}{\partial x} - \frac{imc}{\hbar}\psi_4 = 0.\]

The first and fourth equations are coupled only to each other, and can be written as:

\[\frac{1}{c}\sigma_0\frac{\partial\psi_a}{\partial t} + \sigma_x\frac{\partial\psi_a}{\partial x} + \frac{imc}{\hbar}\sigma_z\psi_a = 0,\]

where

\[
\sigma_0 \equiv
\left(
\begin{array}{ccc}
1 & 0\\
0 & 1\\
\end{array}
\right),
\]

\[
\psi_a\equiv
\left(
\begin{array}{c}
\psi_1 \\
\psi_4 \\
\end{array}
\right).
\]

The second and third equations are coupled only to each other, and can be written as:

\[\frac{1}{c}\sigma_0\frac{\partial\psi_b}{\partial t} + \sigma_x\frac{\partial\psi_b}{\partial x} + \frac{imc}{\hbar}\sigma_z\psi_b = 0,\]

where

\[
\psi_b\equiv
\left(
\begin{array}{c}
\psi_2 \\
\psi_3 \\
\end{array}
\right).
\]

These two equations are the same equation, so we can eliminate one to obtain the (1+1)D DE:

\[
\frac{1}{c}\sigma_0\frac{\partial\psi}{\partial t} + \sigma_x\frac{\partial\psi}{\partial x} + \frac{imc}{\hbar}\sigma_z\psi=0,
\]

where
\[
\psi \equiv
\left(
\begin{array}{c}
\psi_1 \\
\psi_2 \\
\end{array}
\right).
\]
 
\end{document}